\documentclass[aps,pra,superscriptaddress,preprint,amsmath,amssymb]{revtex4}
\usepackage{bbm}
\usepackage{graphicx}
\usepackage{epstopdf}

\newcommand{\Op}[1]{\boldsymbol{\mathsf{\hat{#1}}}}

\def\cm{\mbox{cm$^{-1}$}}
\def\hm2{\frac{\hbar^2}{2m}}
\def\hmi{\frac{\hbar}{2m i}}

\def\cm{\mbox{cm$^{-1}$}}

\def\E{E_{\textrm{pulse}}}
\def\gts{\mbox{a$^3\Sigma_{\textrm{u}}^+$}}
\def\zerog{\mbox{0$_g^-$(6S+6P$_{3/2}$)}}
\def\Rmin{\mbox{$R_{\textrm{min}}$ }}
\def\Rmax{\mbox{$R_{\textrm{max}}$ }}
\def\vmin{\mbox{$v_{\textrm{min}}$ }}
\def\vmax{\mbox{$v_{\textrm{max}}$ }}
\begin{document}

\title{The dynamical hole in ultrafast photoassociation: analysis of the compression effect}
\author{Eliane Luc-Koenig}
\affiliation{Laboratoire Aim\'e Cotton, CNRS, B\^{a}t. 505, Campus d' Orsay,
91405 Orsay Cedex, France}
\author{Fran\c{c}oise Masnou-Seeuws}
\affiliation{Laboratoire Aim\'e Cotton, CNRS, B\^{a}t. 505, Campus d' Orsay,
91405 Orsay Cedex, France}
\author{Ronnie Kosloff}
\affiliation{Department of Physical Chemistry and
The Fritz Haber Research Center, 
The Hebrew University, Jerusalem 91904, Israel}

\begin{abstract}
Photoassociation of a pair of cooled atoms 
by  excitation with a short chirped laser pulse creates a dynamical hole
in the initial  continuum wavefunction.  This hole is manifested by a void in the pair wavefunction and a momentum kick. Photoassociation into loosely bound levels 
of the external well in Cs$_2$ \zerog \ is considered as a case study. After the pulse, the free evolution of the ground triplet state \gts wavepacket  is analyzed.  Due to a negative momentum kick, motion to small distances is manifested and a compression effect is pointed out, markedly increasing  the density of  atom pairs at short distance. A consequence of the hole is the redistribution of the vibrational population in the \gts state, with population of the last bound level and creation of pairs of hot atoms.  The physical interpretation makes use of the time dependence of  the probability current and population on each channel to understand the role of the  parameters of the photoassociation pulse. By varying such parameters, optimization of the compression effect in the ground state wavepacket is demonstrated.  Due to an increase of the short range density probability by more than two orders of magnitude, we predict important photoassociation rates into deeply bound levels of the excited state by a second  pulse, red-detuned relative to the first one and conveniently delayed. 
\end{abstract}

\maketitle

\section{Introduction}
\label{sec:intro}
One of the main outcomes of  excitation of a molecular system with a short laser pulse
is a "hole" in the wavefunction of the initial ground state  \cite{k108,guy97}. 
The impulsivee limit has been extended 
to the problem of photoassociation of ultracold atoms Ref. \cite{luc2004a}. 
Due to the ultralow collision energy this limit can be applied to a well-defined initial
stationary collision state. The emphasis of that
study was on optimizing the photoassociation process and the creation of
stable ultracold molecules.  We should complete the story and follow the dynamics 
of the "hole" carved out of the initial continuum wavefunction. We have already 
identified the signatures of this "hole": 
\begin{itemize}
\item{1) a void in the pair wave function of the initial state}
\item{2) a momentum kick, initiating a  motion of the hole}
\item{3) creation  of stable molecules , via population of the last bound levels of the ground electronic state}.
\item{4) creation of pairs of hot atoms, via population of continuum levels with higher energy }
\end{itemize}
The aim of the present paper is to analyze in detail the hole created in the ground state continuum wave function, during the photoassociation pulse and after it. In Section \ref{sec:example}, we display results of numerical calculations on photoassociation of cesium with chirped laser pulses, showing the time-evolution of the wave packet in the initial state. We identify the dynamical hole, its motion, and the compression of the initial wavefunction, increasing the probability density at short internuclear distances. Possible applications of this compression effect to photoassociation with a second pulse are suggested. In Section \ref{sec:tools} we present theoretical tools for the analysis of a dynamical hole through the expectation value of the current density vector. In Section \ref{sec:analysis} we make use of such tools to analyze the numerical results of Section \ref{sec:example} and their sensitivity to the pulse parameters (energy, duration, central frequency, sign a
 nd value of the chirp parameter). In Section \ref{sec:control} we discuss how to control the formation of a dynamical hole and the compression effect, and to optimize photoassociation into vibrational levels with low $v$ values.

\section{Phenomenological description of the dynamical hole and of the compression effect: numerical calculations in case of cesium photoassociation.}
\label{sec:example}
\subsection{The physical system}
\begin {figure}[htbp]
\centering
  \includegraphics[width=0.7\textwidth,angle=270]{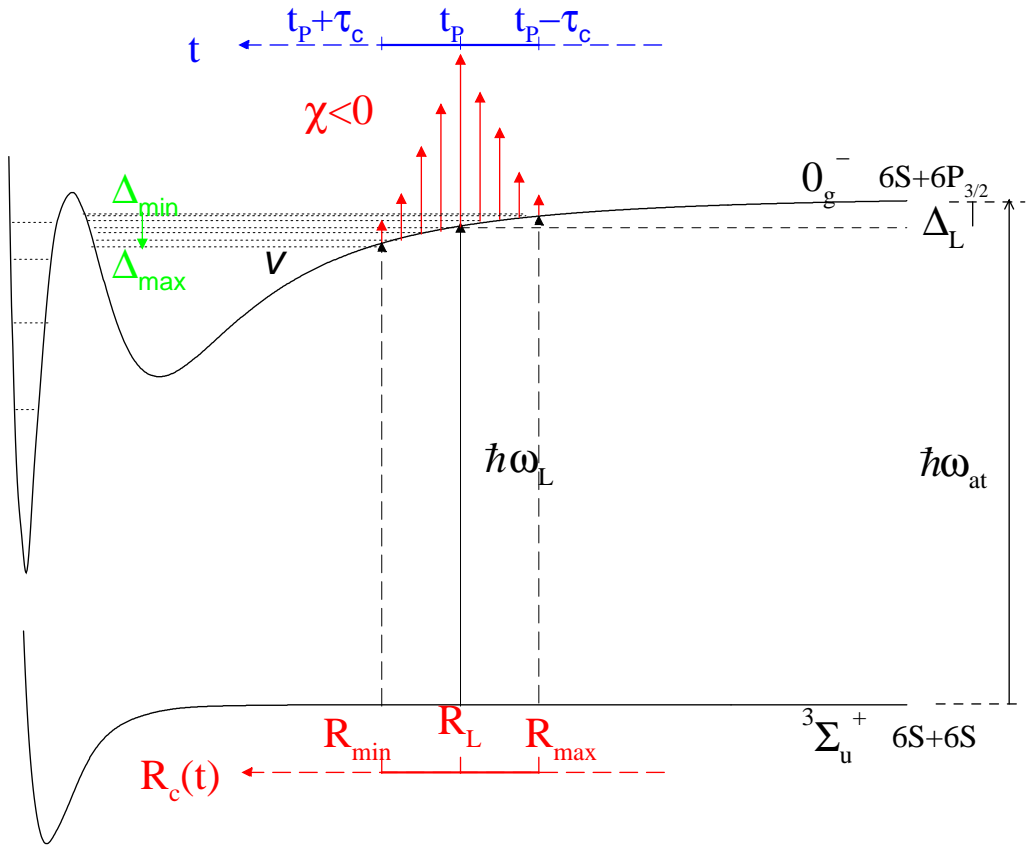}
  \caption{(color online)The potential curves for the lower triplet state and the double well $0_g^-$(P)$_{3/2}$ excited state. Also indicated are the photoassociation windows in the time $t$ (blue), energy $\Delta$(green) and crossing distance $R_C(t)$ (red) domains in case of a pulse with negative chirp.}
  \label{fig:potentiels}
  \end{figure}

The photoassociation reaction
\begin{equation}
2 \mathrm{Cs}(6S,F=4)+ \hbar(\omega(t)) 
 \rightarrow \mathrm{Cs}_{2}(0_g^-(6S+6P_{3/2});v,J)
\label{eq:photoass}
\end{equation}
is employed as a primary example.
In this reaction two ultracold cesium atoms absorb a photon red detuned from the resonance line to form a bound level $(v,J)$ in the outer well of the excited potential curve $0_g^-(6S+6P_{3/2})$. The potential curves have been described elsewhere \cite{luc2004a} and are displayed in Fig. \ref{fig:potentiels}. We shall use vibrational numbering for the double-well potential : the external well supports at least 227 levels, from $v=25$ up to $v=256$, the levels $v$=33, 45, 57, 93, 127 belonging to the inner well . In the chosen example the laser pulse is 
a Gaussian chirped pulse of energy $\E$,
centered at time $t_P$, with a frequency 
\begin{equation}
  \omega(t) = \omega_L + \chi\cdot(t-t_P)=\frac{d}{dt}(\omega_L t + \phi(t)) \:,
  \label{eq:chirp}
\end{equation}
that varies linearly around the carrier frequency $\omega_L$. In Eq. (\ref{eq:chirp}), we have introduced the phase $\phi(t)$ of the electric field. 
The laser is red-detuned from the atomic D$_2$ line at $\omega_{\textrm{at}}$
by $\Delta_L=\hbar(\omega_{\textrm{at}}-\omega_L)$,
$\chi$ is the linear chirp rate in the time domain.
The spectral bandwidth $\delta\omega$, defined as the FWHM of the intensity profile is related to the duration $\tau_L$ of the transform limited pulse with the same bandwidth by
$\delta\omega=4\ln2 / \tau_L \approx 14.7~\cm / \tau_L[\textrm{in~ps}]$.
The instantaneous intensity of the pulse involves a Gaussian envelope,
\begin{align}
  \!\!\!I(t) &\!=\frac{\!E_{\mathrm{pulse}}}{\sigma\tau_C}
  \sqrt{\frac{4\ln2}{\pi}}
  \exp\!\!\left[ -4\ln2
      \left(\frac{t-t_P}{\tau_C}\right)^2 \right]=I_L[f(t)]^2,
\label{eq:pulse}
\end{align}
with a FWHM equal to $\tau_C$ ($\geq\tau_L$),
the pulse duration~\cite{luc2004b}.
These parameters are related by
$\chi^2\tau_C^4=(4\ln2)^2[(\tau_C/\tau_L)^2-1]$.
For this pulse, 98\% of the energy is delivered in the
{\em time window} $[t_P-\tau_C,t_P+\tau_C]$
over the illuminated area $\sigma$~\cite{luc2004a}.
During this time, the instantaneous laser frequency is resonant with
the  excited levels with a binding energy in the range
$[\Delta_L-\hbar|\chi|\tau_C,\Delta_L+\hbar|\chi|\tau_C]$,
which defines a {\em photoassociation window in the energy} domain.  Assuming the reflexion principle to be valid, this window  can be translated into a {\it {window in the $R-$domain}}, considering the outer classical turning points of the set of levels. As discussed in \cite{koch2006c}, for large enough values of the detuning, the photoassociation window in the $R-$domain is located in a region where the nodal structure of this initial wavefunction is  energy-independent in the range defined by the thermal distribution, so that  thermal averaging is straightforward. When this is not the case, numerical averaging is possible \cite{koch2006c}.
\subsection{The two-channel coupled equations}

The vibration dynamics in the ground  and the excited electronic states is described by
the time-dependent Schr\"odinger equation
\begin{equation}
\mbox{$\mathbf{\hat{H}}$} \mbox{$\mathbf{\Psi(t)}$} = (\mbox{$\mathbf{\hat{H}_{mol}}$} +
\mbox{$\mathbf{\hat{W}(t)}$}) \mbox{$\mathbf{\Psi(t)}$}= i \hbar
\frac{\partial}{\partial t} \mbox{$\mathbf{\Psi(t)}$}. 
\label{eq:tse}
\end{equation}
where $\mathbf{\Psi(t)}$ is a two-component wavefunction describing the relative motion of the nuclei in the two channels.
The molecular Hamiltonian $\mathbf{\hat{H}_{mol}} = \mathbf{\hat{T}} + \mathbf{\hat{V}_{el}}$  is the sum of the kinetic energy operator $\mathbf{\hat{T}}$  and electronic potential energy operator $\mathbf{\hat{V}_{el}}$ , with components $V_{\mathrm{ground}}$ and $V_{\mathrm{exc}}$ on the ground and excited surface respectively . The coupling term is written in the dipole approximation:
\begin{eqnarray}
\mbox{$\mathbf{\hat{W}}$}= -\vec{D}(\vec{R} ) \cdot \vec{e_L}\mbox{{{\cal {E}}}(t)}\\
 {\cal{E}}(t)={\cal{E}}_0 \cos(\omega_L t + \Phi(t)) \nonumber
\label{eq:hdip}
\end{eqnarray}
involving the transition dipole moment of the dimer $\vec{D}(\vec{R})$ 
and the electric field defined  by  a polarization vector $\vec{e_L}$ 
assumed to be constant and by an amplitude ${\cal{E}}(t)$.

The explicit temporal dependence of the Hamiltonian $\mathbf{\hat{H}}$ 
is eliminated in the framework of the rotating wave approximation. In the present paper, this approximation considers  the central laser frequency $\omega_L$, 
multiplying the radial wavefunction for the nuclear motion in the ground 
and the excited states by exp$(-i\omega_L t/2)$ and exp$(+i \omega_L t/2)$ 
respectively, and neglecting the high frequency component in the coupling term. 
This allows to write the radial coupled equations as
\begin{eqnarray} 
&&i\hbar\frac{\partial}{\partial t}\left(\begin{array}{c}
 \Psi_g(R,t)\\
\Psi_e(R,t)
 \end{array}\right) \nonumber\\
&&=
 \left(\begin{array}{lc}
 {\bf \hat T} + V_g(R)  & 
W^*_L \exp(i\phi) \\
 W_L \exp(-i\phi)& 
 {\bf \hat T} +  V_e(R)
 \end{array} \right) 
 \left( \begin{array}{c}
 \Psi_g(R,t)\\
\Psi_e(R,t) 
 \end{array}\right),
 \label{eq:eqcpl}
 \end{eqnarray}
where the potentials are now crossing 
\begin{equation}                                      
V_g(R)= V_{\mathrm{ground}}+\hbar \omega_L/2,
 V_e(R)=V_{\mathrm{exc}}-\hbar\omega_L/2.
\label{eq:dressed_pot}
\end{equation}
and the coupling term is defined as  
\begin{equation}
W_L(R,t) \exp(-i\phi(t))=-\frac{1}{2}\vec{D}(\vec{R} ) \cdot \vec{e_L}{\cal {E}}_0 f(t)exp(-\frac{i}{2}\chi(t-t_P)^2)
\label{eq:W} 
\end{equation}
In the following calculations, the $R$-dependence of the dipole moment was not considered.

\subsection{Results of the time-dependent calculations}
\label{ssec:calculs}
Time-dependent calculations were performed by integrating Eq.(\ref{eq:eqcpl}), the numerical method have been  described previously \cite{luc2004a,luc2004b}.
Due to the low collision energy considered only $s-$wave scattering has to be accounted. The initial state is represented by a stationary collision wave function, $\Phi_g(R,t_{\mathrm{init}})$ as illustrated in the upper panel of  Fig. \ref{fig:hole}, corresponding to a collision energy $E/k_B$=54 $\mu$K. For a scattering length $a \sim $ 300 a$_0$, the last node in the inner region (not considering the pure sine behaviour of the wavefunction for a non zero energy), hereafter referred-to as "last node" is located at $R_a$ = 310 a$_0$($\sim a$), and the last but one at $R_{\mathrm {last1}}$= 82.5 a$_0$. In an energy range close to threshold, all nodes with $R \le R_{\mathrm {last1}}$ are  energy-independent.
The cold atoms are illuminated by a chirped laser pulse , referred to as ${\cal P}^{122}$ in Refs \cite{luc2004b,koch2006c,koch2006b}, with central detuning $\Delta_L$ = 0.675 \cm, duration (FWHM) $\tau_C=110$ps, stretching factor $f_P=\tau_C/\tau_L$= 1.91 associated with a transform limited pulse of duration $\tau_L$=57.5ps and maximum intensity $I_L$= 120 kW cm$^{-2}$. The central frequency $\omega_L$ is resonant with the level $v$=153, for which  the outer turning point is located at $R_L=$~148.5~a$_0$ ( corresponding to a maximum of the wavefunction, beween $R_{\mathrm {last1}}$ and $R_a$). For a negative chirp, the photoassociation window spans the levels $v$= \vmax = 159 till  \vmin =149,  bound by 0.456 to  0.869  \cm , with a classical vibrational period $T_\mathrm{vib}$ from 1095 till  635 ps. In the $R-$domain, the window extends from  \Rmin = 135 till  \Rmax = 176 a$_0$.

The conditions for an impulsive approach are that the pulse duration is shorter than the vibrational periods (see Ref. \cite{luc2004a}). At the intensity considered,  during the pulse many levels are populated, but after the pulse only the levels between \vmin and \vmax  remain populated \cite{luc2004a}. 

\subsection{Momentum kick on the ground $^3\Sigma_u^+$ electronic state}

The generation of the depletion hole on the ground $a{^3\Sigma_u^+}$ electronic state  can be understood by employing a phase amplitude representation both for the ground and excited state functions:
\begin{equation}
 \Psi_{g,e}(R,t)=A_{g,e}(R,t)\exp[\frac{iS_{g,e}(R,t)}{\hbar}]=A_{g,e}(R,t)\exp[\frac{i\int^R p_{g,e}(R,t)dR}{\hbar}],
 \label{eq:phaseamp}
\end{equation}
where $p_g(R,t)$ can be interpreted as the local momentum  in the ground state \cite{messiah,schiff}.
The evolution of the amplitude $A_g(R,t)$, displayed in Fig. \ref{fig:hole}, suggests the following remarks: 
\begin {figure}[htbp]
\centering
  \includegraphics[width=0.8\textwidth,clip=true]{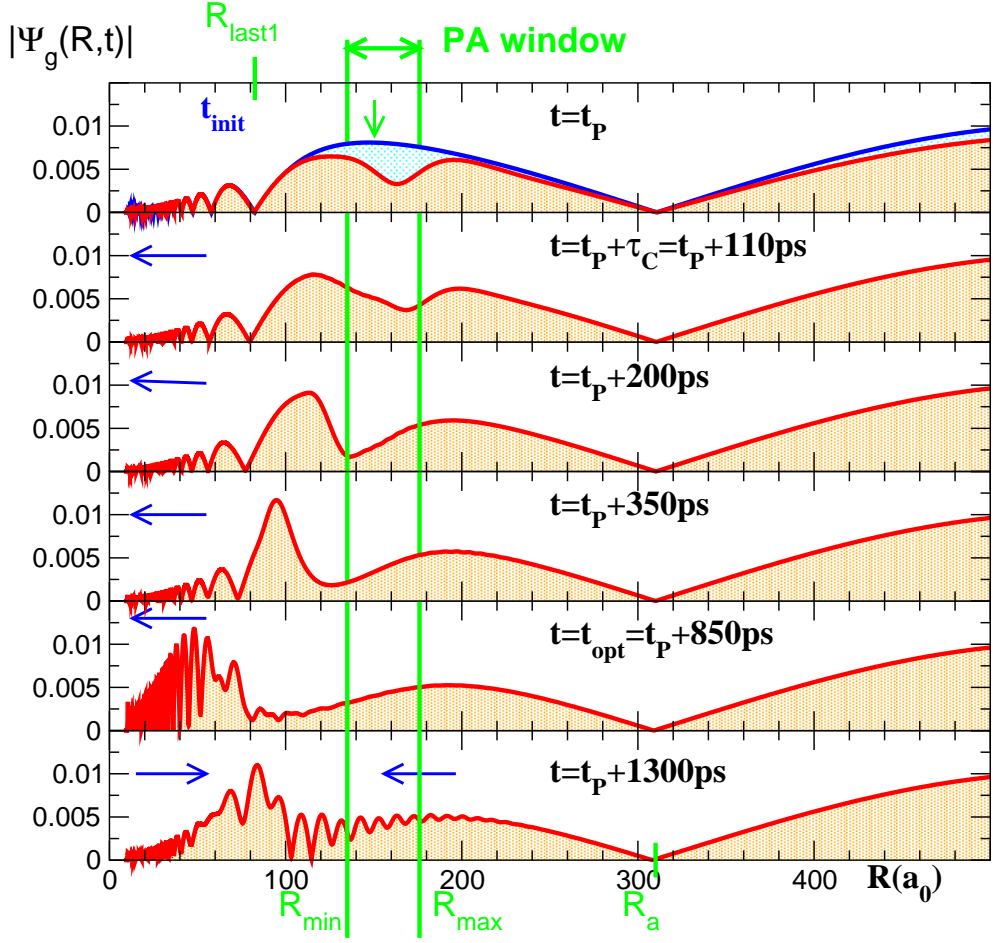}
  \caption{(color online) Upper panel (blue line):  amplitude $A_g(R,t=t_{\mathrm{init}})$ of the initial scattering stationary wavefunction  for a pair of ground state cesium atoms colliding with energy $E=k_BT, T=54 \mu K$,  and zero angular momentum, in the lack of any electromagnetic field.Note the position of the "last node" at $R_a$ and the last but one at $R_{last1}$. Upper panel(red  line) :  amplitude $A_g(R,t=t_P)$ at the maximum of the pulse $t=t_P$  : the pulse has carved out a hole in the initial wavefunction. This hole is indeed located in the region $[R_{\textrm{min}}, R_{\textrm{max}}]$ of the photoassociation window, delimited by the two vertical lines.
Next panels : after the pulse, the $\Psi_g(R,t)$ wavepacket moves inwards : note the increase of the maximum value of the amplitude in the inner region.
At $t-t_{\mathrm P}  \sim$ 850 ps, the wavepacket has been partly  reflected by  the ${^3\Sigma_u^+}$ inner wall located at $R \sim 10a_0$. Note that in the excited state the classical vibrational half-period for the photoassociated levels is in the $\sim$ [300,600] ps range}
\label{fig:hole}
\end{figure}

\begin{itemize}
\item Before the pulse, we see in the upper panel the initial state stationary collision wavefunction $|\Psi_{g}(R, t_{\mathrm{init}})|$ described above, and computed as a unity normalized eigenstate in a box of length $L$=19 250 a$_0$. In order to optimize the population transfer, we have chosen a detuning such that the photoassociation window [135 , 176 a$_0$] is located in the region of the "last" maximum of $|\Psi_{g}(R,t=0)|$.
\item  
The laser pulse, maximum at $t=t_P$, carves out  a depletion hole in the initial wavepacket.  The location of this hole is consistent with transfer of population to the levels $v$=149 to 159, with outer turning points from 135 to 176 a$_0$, located within the photoassociation window. Note that since $\int_{135}^{176} |\Psi_{g}(R,t=0)|^2 \mathrm{d}R$ = 0.0025, only a very small fraction of the probability density in the initial state is transferred to the excited state. This very small value is due to the fact that we use a very large box. Calculations using energy-normalized initial state have been reported elsewhere \cite{luc2004a,luc2004b}, as well as determination of the absolute value of photoassociated molecules \cite{koch2006c}.  
\item  After the pulse, this hole starts moving to smaller internuclear distances, due to a negative "momentum kick" (see Ref. \cite{k108,guy97}  and Sec.\ref{sec:tools} below). Whereas the motion is taking place in a region where both  $V_g$ and $\frac{\mathrm{d}V_g}{\mathrm{d}R} $ are negligible, the classical velocity is quite impressive : for instance, the  new maximum created on the left side of the hole and initially located at $R$=110 a$_0$ moves to $R$=90 a$_0$ with a velocity $\sim$ 4.2 m s$^{-1}$, two orders of magnitude larger that the classical velocity of the colliding atoms.
\item The motion is restricted to a localized part of the wavepacket : during the time-period considered here, the position of the "last" node at $R_a$= 314 $a_0$, and the wavefunction at distances larger than $R_a$, are not modified. 
\item In contrast, we observe  an increase of the probability density at  distances shorter than the hole position, hereafter referred to as "compression" of the wavepacket. The amplitude of the secondary maximum created on the left side of the hole increases and has grown by a factor of 2 when it reaches   $R \sim$90 a$_0$. Therefore, at times  $t-t_P \ge $350 ps, (the optimal value being 850 ps ), the wavepacket is well adapted to photoassociation with a second pulse, red-detuned relative to the first one, populating in the excited state vibrational levels  with an outer turning point around 90 a$_0$. 
 
\item 
At $t-t_P$=850 ps, the wave packet is reflected by the inner wall of the \gts  potential, and starts moving outwards. Note that  in the excited state the classical vibrational half-period varies from 318 for $v_{min}$ to 548 ps  for  $v_{max}$. 
\item We may therefore conclude that after the pulse the motion of the wavepacket in the ground state is governed by a timescale similar to the vibrational motion in the photoassociated levels of the excited state.
\end{itemize}

\begin{figure}[htbp]
\centering
\includegraphics[angle=270,width=0.9\textwidth,clip=true]{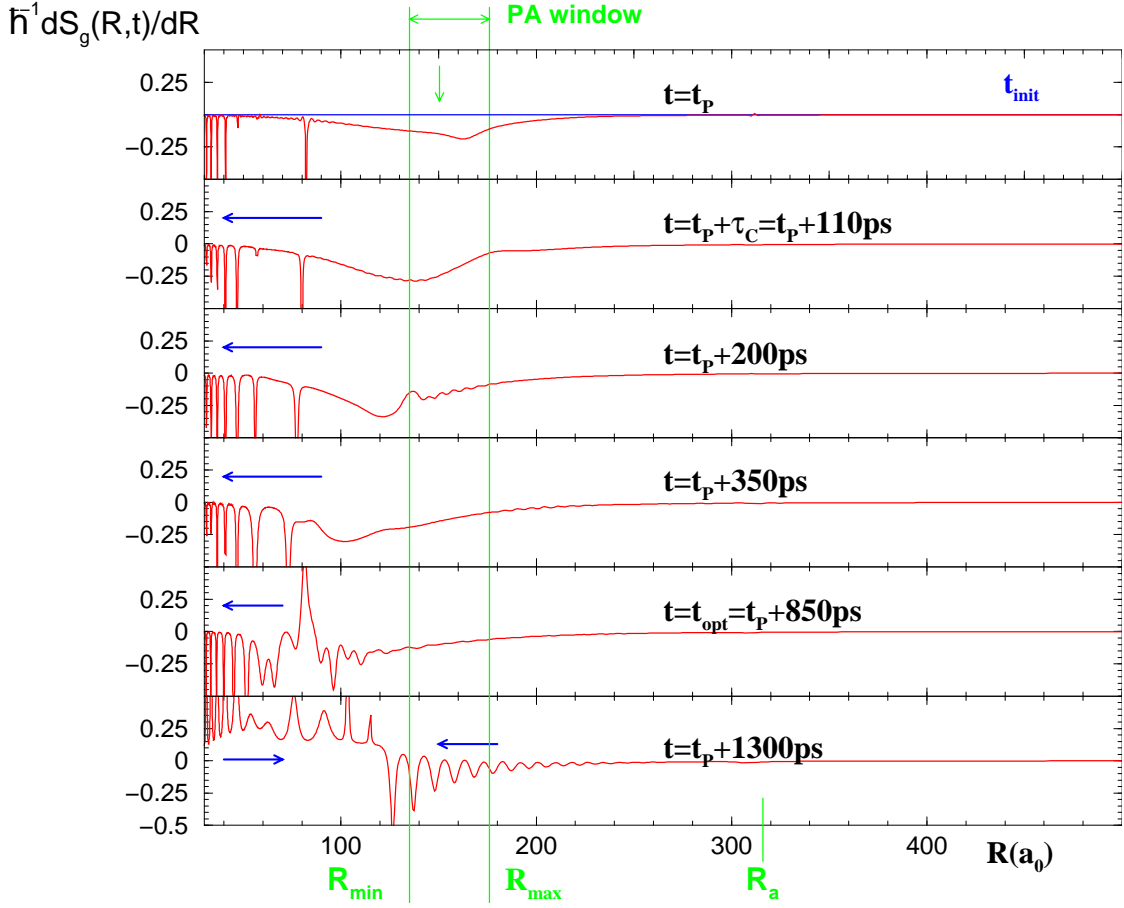}
\caption{(Color online)  Same as Fig. \ref{fig:hole} for the variation of the semiclassical momentum $p_g(R,t)=\partial S_g(R,t)/ \partial R$, where $S_g(R,t)/\hbar$ is the phase of the wavefunction in the ground state (see Eq. (\ref{eq:phaseamp}) in text) . Just after the pulse, the value $\sim$ 0.20 (a$_0)^{-1}$ of the momentum created in the photoassociation window is consistent with a classical velocity of 3.6 m s$^{-1}$. As expected the momentum is increased in the short range region where the potential $V_g(R)$ is not negligible.}
\label{fig:phase}
\end{figure}

We next display in Fig. \ref{fig:phase} several snapshots for the variation of the phase derivative $\frac{\partial S(R,t)}{\hbar \partial R}=p_g(R,t)/\hbar$ (see Eq. (\ref{eq:phaseamp})), where $p_g(R,t)$ can be interpreted as the semiclassical momentum \cite{messiah}. Just after the pulse, significant momentum is created in the region of the photoassociation window, where the potential and its gradient are negligible. The computed value $\sim$ 0.20 (a$_0)^{-1}$ corresponds to a classical velocity of 3.6 m s$^{-1}$, same order of magnitude as the very rough estimation  of 4.2 m s$^{-1}$ given previously. As expected, a strong momentum increase is observed in the short range region ( $R < 30$ a$_0$), where the potential is strongly attractive and the wavefunction rapidly oscillating. This will be analyzed further in Sec. \ref{sec:tools}.

\subsection{Advantage of the compression effect for photoassociation with a second pulse}

\begin{figure}[htbp]
\includegraphics[width=.7\textwidth]{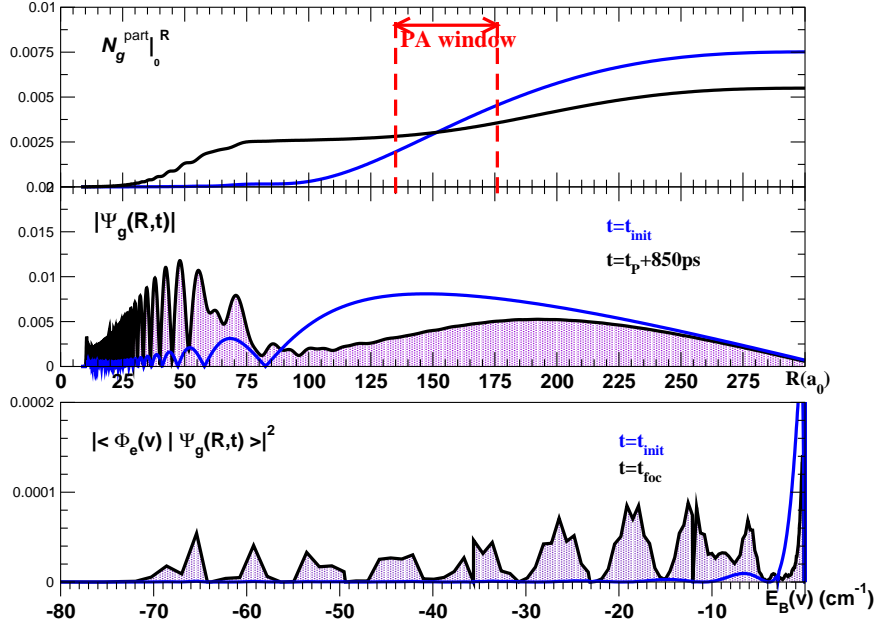}
\caption{ (Color on line)Interest of the compression effect. Intermediate panel : wavefunction $|\Psi_g(R,t=t_{\mathrm{init}})|$  of the initial collision state and ground state wavepacket $|\Psi_g(R,t_{\mathrm{opt}}=t_P+850)|$, 850 ps after the pulse maximum, when the compression effect is maximum. Upper panel: integrated population $\int_0^R |\Psi_g(R,t)|^2$, showing the increase of the probability density at short distances. Lower panel : variation of the overlap $|<\Phi_e(v)|\Psi_g(R,t)>|^2$  with the eigenfunctions for all the  bound vibrational levels $v$ in  the $0_g^-(6S+6P_{3/2}$) excited potential curve, as a function  of the  binding energy $E_B(v)$, for $t=t_{\mathrm{init}}$(black) and $t=t_{\mathrm{opt}}$. Several dips visible in the overlap should be attributed to levels in the inner well ( $v$=33, 45, 57, 73,93, 127), which are indicated in red.}\label{fig:overlap}
\end{figure}

To explore further the possibilities offered by the compression of the wavefunction we have represented in Fig. \ref{fig:overlap} the time-variation of the Franck-Condon overlap between $\Psi_g(R,t)$ and the  wavefunctions  of the bound vibrational  levels in the excited state. 
Almost all levels in the external well of the  $0_g^-(6S+6P_{3/2}$) excited potential curve can be efficiently  populated. The deepest level populated  is $v$=29, $v_{ext}=4$ with a binding energy of 70 \cm, an inner turning point at $R$=30 a$_0$, and a good Franck Condon overlap with the $v''$= 42 and 33 levels of the \gts \ potential, respectively bound by 3.7 and 20.5 \cm. Population of such levels might be interesting for the implementation of the stabilization step.

\subsection{Redistribution of population in the ground state : formation of pairs of hot atoms and of halo molecules}

As a result of the short pulse, the population in the ground state is redistributed. Indeed, the photoassociation reaction (\ref{eq:photoass}) is accompanied  by a redistribution process,
\begin{equation}
2 \mathrm{Cs}(6S,F=4) (E=k_BT)+ \hbar(\omega(t),) 
 \rightarrow 2 \mathrm{Cs}(6S,F=4) (E'=k_BT')
\label{eq:redistrib}
\end{equation}
producing pairs of ground state atoms with different energies ( in particular, pairs of "hot" atoms that may leave the trap) and by an association reaction, 
\begin{equation}
2 \mathrm{Cs}(6S,F=4)+ \hbar(\omega(t)) 
 \rightarrow \mathrm{Cs}_{2}(^3\Sigma_u^+(6S+6S);v'',J ),
\label{eq:last}
\end{equation}
with formation of bound molecules in the last vibrational levels of the \gts  potential.

\begin{figure}[htbp]
\includegraphics[angle=270,width=.7\textwidth]{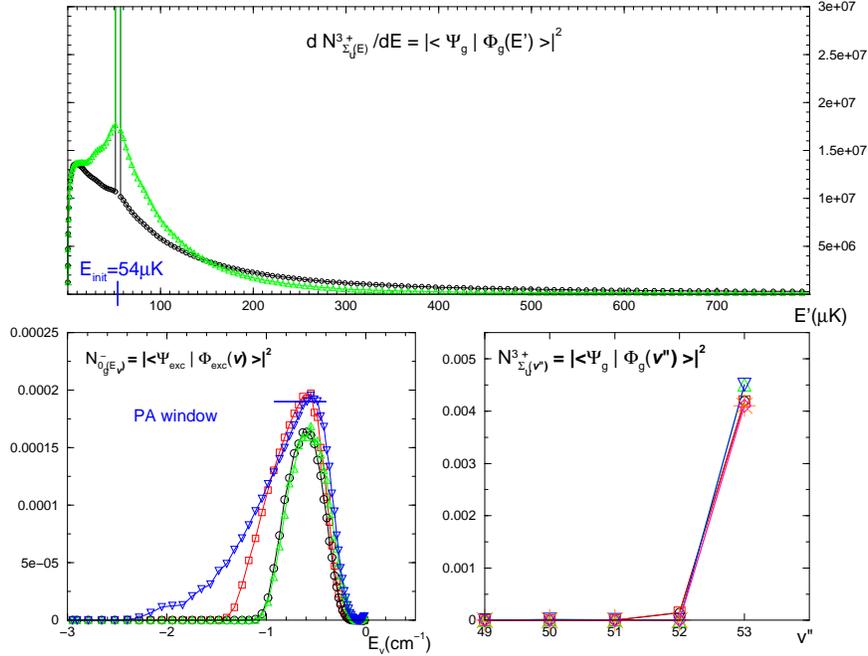}
\caption{(Color on line)Population redistribution in the ground state : analysis of the wavepacket  after illuminating with the pulse $P^{122}_-$, with negative chirp (black circles), and $P^{122}_+$ (red squares). Redistribution after a similar pulse with higher intensity, changing the maximum from $I_L$ to 9$I_L$( green up triangles, $\chi <0$, blue down triangles, $\chi >0$). Upper panel : redistribution in the continuum, as a function of the energy, for an initial state $E_{init}$=54 $\mu$ K. The continuum levels are energy normalized. The results do not depend upon the sign of the chirp. Right lower panel : population transfer to the last bound levels of the ground triplet state, as a function of the vibrational index : only the last level is populated. Left lower panel: for comparison, population transfer to the bound levels of  the excited state, as a function of their binding energy.}
\label{fig:redistrib}
\end{figure}

The redistribution  into the stationary levels $\Phi_g(v'')$ of the ground state potential  is analyzed in Fig.~ \ref{fig:redistrib} as a function of the final collision energy $(E'=k_BT')$  by projecting the wavepacket $\Psi_g(R,t)$ after the pulse on eigenstates (energy-normalized continuum levels and bound levels) of the (${\bf \hat T} + V_g(R)$) Hamiltonian. The calculations are considering a unity-normalized initial state. We have studied the sensitivity to the pulse parameters. The redistribution is independent of the sign of the chirp; in contrast, when the maximum intensity of the pulse $I_L$ (see Eq. (\ref{eq:pulse})) is increased,  the width of the redistribution function is clearly reduced.  Note that the area below the  curve is roughly conserved, changing from 0.0097 to 0.011 when the intensity increases by a factor of 9.  

Population is also transferred efficiently to bound levels of the ground triplet state, primarily to the last bound level $v''$=53, creating a halo molecule. This effect is similar  to the population
of the last bound level by sweeping an optical Feshbach resonance. 
The number of ground state halo molecules (0.004) is the same order of magnitude as the number of
photoassociated molecules (0.002) as can be seen
by comparing the left and right lower panels of Fig. \ref{fig:redistrib}, in accordance with Ref. \cite{luc2004a}. It should be noticed that the population of the photoassociated molecules is spread over more than 30 vibrational levels. 

When the sign of the chirp is changed to a positive value, the population of the last bound level and the redistribution into continuum levels is not affected. In contrast, the photoassociation efficiency is different (see left panel of Fig. \ref{fig:redistrib}). For $\chi<0$, the levels populated in \zerog are within the photoassociation window, and the population  depends weakly upon $I_L$. For $\chi>0$, the distribution extends to levels outside the photoassociation window, and this effect increases with intensity.  As discussed in Ref. \cite{brown2006,wright2007}, for $\chi<0$ the instantaneous frequency "follows" the motion of the wavepacket, so that the population is brought back to the ground state and up again. This phenomenon is labelled "multiple interaction" in Ref. \cite{wright2007}. For $\chi>0$, this recycling effect is much weaker : once a level is populated, the instantaneous frequency is no longer resonant with its energy, and the population remains in this photoassociated level. This recycling effect will be discussed further in the following sections.

\section{The current density vector as a tool for interpretation}
\label{sec:tools}
\subsection{Local current density vector and local momentum}

In order to get better  insight into the hole dynamics one has to follow both the population
depletion as well as  the local value of the momentum induced by the laser field \cite{k108,guy97}.
For the problem of photoassociation, a global quantum analysis of the total changes in population and in momentum due to  the pulse
has been discussed recently in Ref. \cite{kallush2007b}. This formalism, based on Heisenberg equations, is adapted to the implementation of local control theory.\\
In the present work we use a local framework closer in spirit to
the phase amplitude separation Eq. (\ref{eq:phaseamp}) and based on the change in time of
the radial component of the probabilty current density \cite{messiah,schiff}. The analysis is carried out for both the ground  
and excited state components of the wavefunction as defined in Eq. (\ref{eq:eqcpl}),

\begin{equation}
j_{g/e}(R,t)=\hmi [\Psi^*_{g/e}(R,t)\frac{\partial \Psi_{g/e}(R,t)}{\partial R}- \frac{\partial \Psi^*_{g/e}(R,t)}{\partial R}\Psi_{g/e}(R,t)].\\
\end{equation}
The local momentum, defined as semiclassical momentum in Eq. (\ref{eq:phaseamp}), and called "state momentum" in Ref. \cite{brown2006}, is related to the probability current density
\begin{equation}
p_{g/e}(R,t)~~=~~\frac{\partial S_{g/e}}{\partial R}~~=~~\frac{mj_{g/e}(R,t)}{|\Psi_{g/e}(R,t)|^2}=\frac{mj_{g/e}(R,t)}{|A_{g/e}(R,t)|^2}
\end{equation}
where the phase and amplitude have been defined in Eq. (\ref{eq:phaseamp}).\\

The formulation is related to the classical limit of Bohmian dynamics \cite{bohm52}, where one considers on each channel the motion of a fluid of non interacting particles with probability density $|A_{g,e}|^2$, probability current density  $j_{g,e}(R,t)$, driven by a force $F_{g,e}(R)=\frac{dp_{g,e}}{dt}=(\frac{\partial }{\partial t}+\frac{\partial}{\partial R}\frac{\partial R}{\partial t})p_{g,e}(R,t)$. This aspect will be discussed in a forthcoming paper \cite{luc2007b}.\\
The time derivative of the two quantities $mj_g(R,t)$ and $mj_e(R,t)$ reads 
\begin{eqnarray}
m \frac{\partial j_g(R,t)}{\partial t}= F_g(R) |A_g(R,t)|^2 + {\cal{K}}_g(R,t) + {\cal{L}}(R,t) \label{eq:accelg} \\
m \frac{ \partial j_e(R,t)}{\partial t}= F_e(R) |A_e(R,t)|^2 + {\cal{K}}_e(R,t) - {\cal{L}}(R,t),
\label{eq:accele}
\end{eqnarray} 
 where $m$ is the reduced mass, and we have defined
\begin{itemize}
\item the classical force  for each potential
\begin{equation}
F_{g/e}(R)=-\frac{\partial V_{g/e}}{\partial R},
\end{equation}
so that the first term in Eqs.(\ref{eq:accelg},\ref{eq:accele}) depends both upon the classical force and the local probability density $A^2_{g/e}(R,t)$ in the channel considered, defining a density of force $F{g/e}(R)A^2_{g/e}(R,t)$ .
\item a kinetic term due to the non-hermiticity of the local kinetic energy operator
\begin{eqnarray}
{\cal{K}}_{g/e}(R,t) = \frac{\partial}{\partial R}[\frac{\hbar^2}{4 m^2}(\Psi^*_{g/e}(R,t)\frac{\partial^2 \Psi_{g/e}(R,t)}{\partial R^2}+ \frac{\partial^2 \Psi^*_{g/e}(R,t)}{\partial R^2}\Psi_{g/e}(R,t))\nonumber \\ 
- 2(\frac{\partial \Psi^*_{g/e}(R,t)}{\partial R}\frac{\partial \Psi_{g/e}(R,t)}{\partial R})]\
\end{eqnarray}
This complicated term, arising from the fact that the time-variation of $j(R,t)$ results both from the time-variation of the local momentum $p(R,t)$ and of the local probability density $|A(R,t)|^2$, will be discussed further in a forthcoming paper \cite{luc2007b}.
\item a third term due to the coupling by the laser field, which cancels  when the laser is off, and has an opposite sign in the ground and in the excited state.
\begin{equation}
{\cal{L}}(R,t)=\Re [\Psi^*_e(R,t)\frac{\partial}{\partial R}(W_L\exp(-i \phi(t))\Psi_g(R,t))] - \Re [W_L^* \exp\left( i \phi(t)\right)\Psi^*_g(R,t)\frac{\partial \Psi_e(R,t)}{\partial R}]
\end{equation}

It depends upon the coupling (amplitude $W_L$, phase $\phi$, and therefore sign of the chirp) and also upon the phase difference between the wavepackets in the ground and in the excited state. 
Introducing the momentum operator $\Op P=-i\hbar\partial/\partial R$ , one may rewrite
\begin{equation}
{\cal{L}}(R,t)=-\frac{1}{\hbar}[\Im (\Psi^*_e(R,t)\Op P(W_L(t) \exp(-i \phi(t))  \Psi_g(R,t)) \\
-\Im (\Psi^*_g(R,t)(W_L^*(t) \exp(+i \phi(t))\Op P \Psi_e(R,t)),
\end{equation}
which is close to the notation of Ref. \cite{kallush2007b} where it was termed the momentum exchange. 
\end{itemize}
Note that the density of force ${\cal{L}}(R,t)$ is zero if the laser field is zero, of course, but also when the local population either in the ground or the excited state is 0.\\

Another set of equations, complementary to Eqs. (\ref{eq:accelg},\ref{eq:accele})  describes the variation of the local probability density 
\begin{eqnarray}
m \frac{\partial A_g(R,t)^2}{\partial t}=
-\frac{\partial j_g(R,t)}{\partial R}-\frac{2}{\hbar}\Im(\Psi^*_e(R,t) W_L(t) \exp(-i \phi(t)) \Psi_g(R,t)), \label{eq:amplig}\\
m \frac{\partial A_e(R,t)^2}{\partial t}=
-\frac{\partial j_e(R,t)}{\partial R}+\frac{2}{\hbar}\Im(\Psi^*_e(R,t) W_L(t) \exp(-i \phi(t)) \Psi_g(R,t)).
\label{eq:amplie}
\end{eqnarray}
 When the laser is off,  Eqs.(\ref{eq:amplig},\ref{eq:amplie}) are the relations of flux conservation on each channel. During the pulse, transfer of population between the two channels is taking place. 
 
\subsection{Integrated and partially integrated values}

With the concept of the photoassociation window in mind we can differentiate between the
local and global influence of the pulse.
For the global effect, we have considered the integrated values, over the total box of length $L$. We define  for the probabiltiy current densities $j_{g/e}(R,t)$, the position probability densities  $|A_{g/e}(R,t)|^2$, and for their time derivatives, the integrated functions: 
\begin{eqnarray}
mI_{g/e}(t)=\int_0^L m j_{g/e}(R,t)dR\\
m \frac{dI_{g/e}(t)}{dt}=m\int_0^L  \frac{\partial j_{g/e}(R,t)}{\partial t}dR\\
N_{g/e}(t)=\int_0^L |A_{g/e}(R,t)|^2 dR\\
\frac{\mathrm{dN}_{g/e}(t)}{\mathrm{d}t}=\int_0^L  \frac{\partial |A_{g/e}(R,t)|^2}{\partial t}dR
\end{eqnarray}
$mI_{g/e}(t)$ is the current and $N_{g/e}(t)$ is the number of particles. In Ref.\cite{kallush2007b}, similar quantities are obtained directly from the Heisenberg equations. The quantities corresponding to the integrated value of ${\cal{K}}_{g,e}(R,t)$ then disappear. In fact, in the present approach,  
when local time-dependent quantities are integrated over the whole box, the hermiticity of the kinetic energy operator results into  $\int_0^R{\cal{K}}_{g,e}(R,t)dR=0$  and one obtains
\begin{equation}
 m \frac{\mathrm{d}I_{g/e}(t)}{\mathrm{d}t}= \int^L_0 F_{g/e} (R) A^2_{g/e}(R,t)dR \pm \int^L_0 {\cal{L}}(R,t)dR
\label{eq:moyenne}
\end{equation}
so that 
\begin{equation}
m \int\frac{\mathrm{d}(I_g(R,t)+I_e(R,t)}){\mathrm{d} t}~=~\int^L_0 ~(F_g (R) |A_g(R,t)|^2 +F_e (R) |A_e(R,t)|^2)dR
\label{eq:sum} 
\end{equation}
where the term due to laser coupling has disappeared.

However, for the analysis  of the photoassociation effect, where the process is taking place within a photoassociation window, a local view is consistent with partially integrated values. We have therefore considered also the time dependent quantities $I_{g/e}^{\mathrm{part}}(t)\left|^{R_2}_{R_1}\right.$ and $N_{g/e}^{\mathrm{part}}(t)\left|^{R_2}_{R_1}\right.$ defined as integrals 
\begin{eqnarray}
 mI_{g/e}^{\mathrm{part}}(t)\left|^{R_2}_{R_1}\right.~=~m\int_{R_1}^{R_2} j_{g/e}(R,t) dR \\
 N_{g/e}^{\mathrm{part}}(t)\left|^{R_2}_{R_1}\right.=\int_{R_1}^{R_2} A^2_{g/e}(R,t) dR
\label{eqn:partgen}
\end{eqnarray}
taken on a limited range of internuclear distances $[R_1,R_2]$, that could be $[R_{min},R_{max}]$. In the general case,  the influence of the kinetic term $\int^{R_2}_{R_1} {\cal{K}}_{g,e}(R,t)$ will manifest itself. This will be discussed in detail in a forthcoming paper \cite{luc2007b}.\\
 
In the present work   the limited range was chosen as $[0,R_a]$, so that $\int^{R_a}_{0} {\cal{K}}_{g,e}(R,t)$ is negligible.  $R_a$ is the position of the "last" node, which stays fixed  during and after  the ${\cal{P}}^{122}_-$ photoassociation pulse (see Fig. \ref{fig:hole}), as discussed above in Section \ref{ssec:calculs}. The justification for such a choice is that at large distance, where the potentials are negligible and the dipole moment $D(R)$ constant, the wavefunction in the excited state merely reflects the initial wavefunction \cite{luc2004a,luc2004b}. Writing $$\Psi_{g/e}(R,t) \sim \alpha_{g/e}(t)\Psi_{g}(R,t_{\mathrm{init}})$$  where $\alpha_{g/e}(t)$ is a complex number, such that $\alpha_{e}(t\gg \tau_C)=0$, it is clear that there is no $R$-dependence of the phase $S_{g/e}(R)$. The analysis of the photoassociation dynamics can then safely be restricted to the $R<R_a$ range.\\

In the following we shall simplify the notations by writing
\begin{eqnarray}
I_{g/e}^{\mathrm{part}}(t)=I_{g/e}^{\mathrm{part}} (t) \left|^{R_a}_0 \right. \\
\mathrm{N}_{g/e}^{\mathrm{part}}(t)=\mathrm{N}_{g/e}^{\mathrm{part}}(t)\left|^{R_a}_0 \right.
\label{eqn:part}
\end{eqnarray}
\newpage
    \section{Analysis of the results}
\label{sec:analysis}

\subsection{Analysis of the results for the pulse ${\cal{P}}^{122}_-$}

\begin{figure}[htbp]
\includegraphics[angle=270,width=0.8\textwidth]{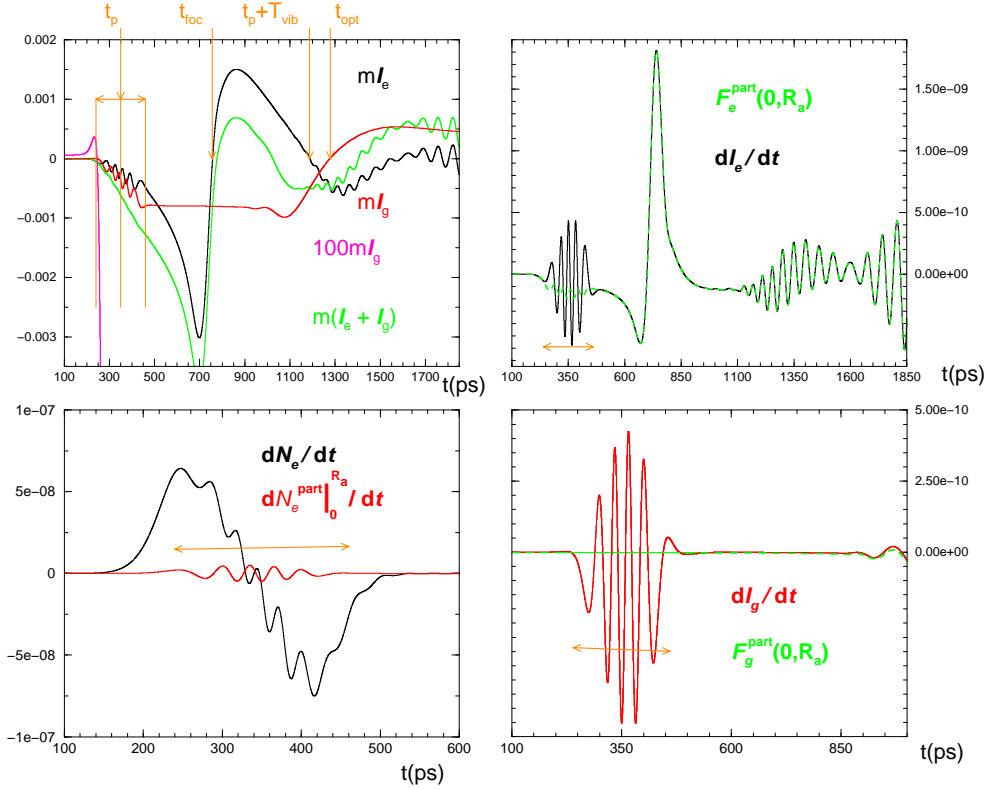}
\caption{ (Color on line) Photoassociation with the pulse ${\cal{P}}^{122}_-$, with negative chirp and small detuning, maximum at $t=t_P$=350 ps: the time-window $[t_P-\tau_C, t_P+\tau_C]$ is indicated by the horizontal arrow. Left column  : in the upper figure  is represented the integrated value   of $m$ times the probability current density  $mI_{g}(R,t)>$ (red line) in the ground and in the excited state $mI_{e}(R,t)>$(black line), and of their sum  as a function of time (green line). Note the Rabi oscillations during the pulse. A small current in the ground state at the beginning of the pulse is drawn as enhanced by  factor 100.  In the lower figure, the time variation of the population is displayed: the calculations have been done by summing over the full range of distances (black lines) and by restricting to $R < R_a~=\sim 314 ~a_0$ (red lines). It is clear that a large amont of population is transferred at large distances during the pulse.  In the right column, the time variation $m \frac{\mathrm{d}I_{g/e}(t)}{\mathrm{d}t}$ (see Eq. \ref{eq:moyenne} in text) in the excited (upper figure) and in the ground state  is compared to the `classical'   term ${\mathrm{F}}^{\mathrm{part}}_{g/e}(0,R_a)$, showing a strong enhancement during the pulse. Note that  the classical force is negligible in the ground state but not in the excited state. After the pulse both curves follow the classical force}
\label{fig:xhineg}
\end{figure}

A comparison of the time-variation of $mI_{g/e}(t)$, $\mathrm{N}_{g/e}(t)$,  and their time derivatives, as well as of the partially integrated quantities $mI^{\mathrm{part}}_{g\/e}(t)$, $N^{\mathrm{part}}_{g/e}(t)$
are reported in Fig.\ref{fig:xhineg}. We have also defined a "partially integrated" force,
\begin{equation}
F^{\mathrm{part}}_{g/e}(0,R_a)=\int_0^{R_a} F_{g/e}(R) A^2_{g/e}(R,t) dR.
\end{equation}

A look at Fig. \ref{fig:xhineg} leads to the following remarks:
\begin{itemize}
\item During the time window $t_p-\tau_C < t< t_p-\tau_C$, there is a large population exchange between the two channels  : in the lower figure of the left pannel,\ $\frac{\mathrm{dN}_{e}(t)} {\mathrm{d}t}=~-\frac{\mathrm{dN}_{g}(t)}{\mathrm{d}t}$ displays  a variation roughly identical to the shape of the pulse, apart from small oscillations. The range of distances $R>$ 314 contributes very much to this population transfer, and we see that the variation of the partial population, $\frac{\mathrm{dN}^{\mathrm{part}}_{e}(t)}{\mathrm{d}t}$, is much smaller than $\frac{\mathrm{dN}_{e}(t)} {\mathrm{d}t}$.
\item After the pulse, the population transfer is restricted to the photoassociation window, as displayed below in Fig. \ref{fig:detail}.
\item In contrast, there is no momentum exchange at large distances, and we find  $I_{g/e}(t) \sim I_{g/e}^{\mathrm{part}}(t)$ . Indeed, in the region  $R > R_a$  the phase  is $R-$independent on both channels($\frac{\partial S_{g,e}(R,t)}{\partial R}=0)$.
\item At the very beginning of the pulse, for times $t < t_p-\tau_C$, when, due to low laser intensity the (small) population transfer is not adiabatic,  the ground state wavepacket gains a positive momentum . The total momentum $I_g + I_e$ is conserved, the contribution of the force being negligible \cite{k108}. 
\item During the time window [$t_p-\tau_C$,$t_p+\tau_C$], there is an exchange of probability current  between the ground and  the excited state. The two currents $I_g$ and $I_e$ are oscillating with opposite phase, in agreement with Eqs.(\ref{eq:moyenne}). Therefore the wavepacket created in the excited state  moves locally back and forth during the pulse.
\item The sum of the probability current on the two channels is not oscillating, in agreement with Eq.(\ref{eq:sum}). Due to the large acceleration  in the excited potential (see the upper figure in the right pannel), this sum current is smoothly decreasing. 
\item For $t_p-\tau_C<t<t_P$ the two channels are equally sharing the classical acceleration due to the $R^{-3}$ asymptotic behaviour of the \zerog potential, so that  $I_g(t)$ and $I_e(t)$ are oscillating with opposite phase below and above their mean value.  
\item For $t_P<t<t_p+\tau_C$ the current density gain in the ground state becomes  larger than in the excited state : the acceleration is no longer equally shared between the two channels. 
\item During the pulse,  the time derivatives $ m \frac{\mathrm{d}I_{g/e}(t)}{\mathrm{d}t}$ are much larger than the classical forces $\int^L_0 F_{g/e} (R) A^2_{g/e}(R,t)dR $ in Eqs. (\ref{eq:moyenne}). The photoassociation dynamics is dominated by  the laser-induced force arising from the term  $\int^{R_a}_0 {\cal{L}}(R,t)dR$. 
\item After the pulse, the inner part of the wavepacket in the excited state is accelerated towards short distances, due the classical acceleration. The pulse ${\cal{P}}^{122}_-$ has been optimized for focussing this vibrational wave packet half a vibrational period after the pulse maximum \cite{luc2004a} : for the levels considered, 318 ps $\le T_{vib}(v)/2 \le$ 550 ps. The  reflexion on the inner wall occurs during a very short time, in the range $t_P-\tau_C +T_{vib}(v_{max})/2, t_P+\tau_C +T_{vib}(v_{min})/2$, $i.e.$ [778,790] ps. In Fig. \ref{fig:xhineg} we may see a sudden change of the sign of $I_e$ at $(t=t_{\mathrm{loc}}\sim $ 784 ps. 
\item In contrast, as can be seen in the lower figure of the right pannel, the classical acceleration is zero in the ground state.  After the pulse, $I_g $ remains constant during a long time delay of $\sim$600 ps. Since the population remains constant, this corresponds to a constant velocity. The part of the wavepacket that is located at $R<R_a$ is moving to short distances with constant velocity for a duration of $\sim$600 ps, in a region where there is no potential gradient. We see then the evidence for a negative momentum kick due to the laser pulse.
\item For  $t-t_P> \tau_C + $600 ps, the wavepacket in the ground state reaches distances where the \gts potential is no longer negligible : it is accelerated at $t$> 1000 ps, then partly reflected by the inner wall at $t\sim$ 1100 ps. At $t=t_{\mathrm{opt}}= 1300$ ps ($t-t_P$= 950 ps), the mean velocity vanishes, which corresponds to the optimal time for reexcitation by a second pulse ( see Fig. \ref{fig:overlap}).

\end{itemize} 
The oscillations which appear during the pulse in the time variation of the populations or of the  current  are analyzed in more detail in Fig. \ref{fig:detail}. These Rabi oscillations are induced by the laser field : indeed, for the ${\cal{P}}^{122}_-$ pulse the  Rabi frequency at resonance, evaluated as $\Omega(t)= \frac{1}{\hbar}|W_L|f(t)$) satisfies $$\int_{t_P-\tau_C}^{t_P+\tau_C} \Omega(R_C,t) dt \sim 5.3 \Pi,$$, which is the same order of magnitude as  the number of  oscillations visible in the figure. \\
\begin{itemize}
\item The $\Pi$ phase difference between the oscillations of the populations $N_{g}$ and $N_{e}$ results from the conservation of population .
\item In a given channel, the  $\Pi$ phase difference between the oscillations of the probability current $I_{g}$ and $I_{e}$ results from the  $\pm \int^L_0 {\cal{L}}(R,t)dR$ laser coupling term in Eq. (\ref{eq:moyenne}).\\ 

\item In each channel, the $\Pi$ phase difference between the time variation of $I$ and $N$ points out that both in Eqs.(\ref{eq:accelg},\ref{eq:accele}) and in Eqs.(\ref{eq:amplig},\ref{eq:amplie}) ,the laser coupling term dominates, so that the change in the current is mainly caused by the change of population.  In other words, the exchange of population between the two channels induces the momentum kick in the ground state:  the population transferred back to the ground state causes a negative momentum kick. It is the Rabi cycling phenomenon, and the efficiency in recycling population that seems the key factor in inducing a motion of the wavepacket towards short distances.
\end{itemize}
\begin{figure}[htbp]
\begin{center}
\includegraphics[angle=270,width=0.7\textwidth]{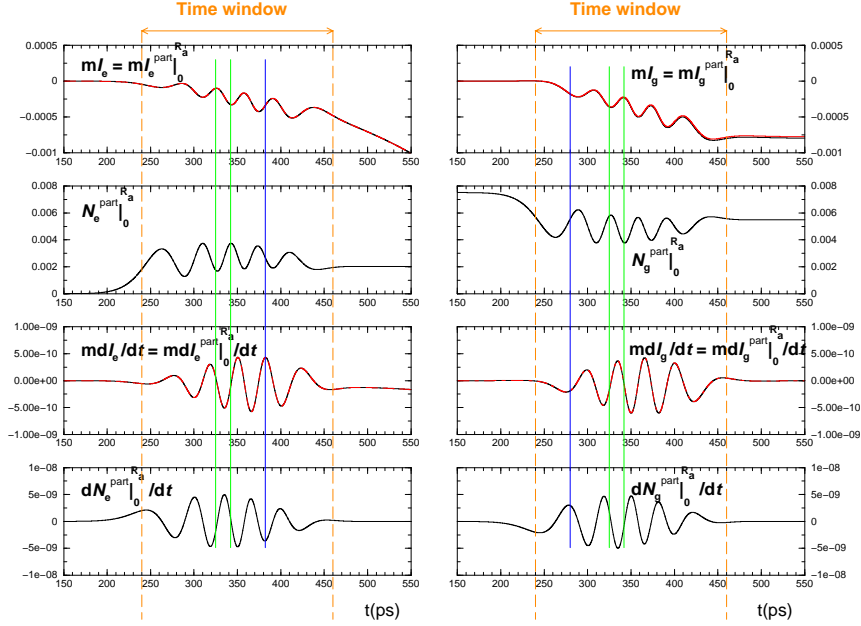}
\caption{Evidence for a momentum kick in the ground state. The laser pulse is P$^{122}$ maximum at time $t=$350 ps. Top figures : Time-variation of the probability current  $I_{e}(t)=I_{e}^{\mathrm{part}}(t)$ (left) and $I_{g}(t)=I_{g}^{\mathrm{part}}(t)=$ (right) on the two channels . Middle pannel : time variation of the partial populations  $\mathrm{N}_{g}^{\mathrm{part}}(t)=$ and $\mathrm{N}_{e}^{\mathrm{part}}(t)=$ in the range $[0,R=314]$ of internuclear distances. Lower pannels :  time-derivatives of the same quantities.  The vertical lines are drawn to emphasize the phase of the Rabi oscillations for each quantity. The pulse is present during the time window [240,460]ps. After the pulse, whereas the  acceleration in the ground state is negligible ($ dI_{g}(t)/dt$=0), the average momentum  is negative and constant, manifesting the momentum kick in the ground state.}
\end{center}
\label{fig:detail}
\end{figure}

\begin{figure}[htbp]
\includegraphics[angle=270,width=0.9\textwidth]{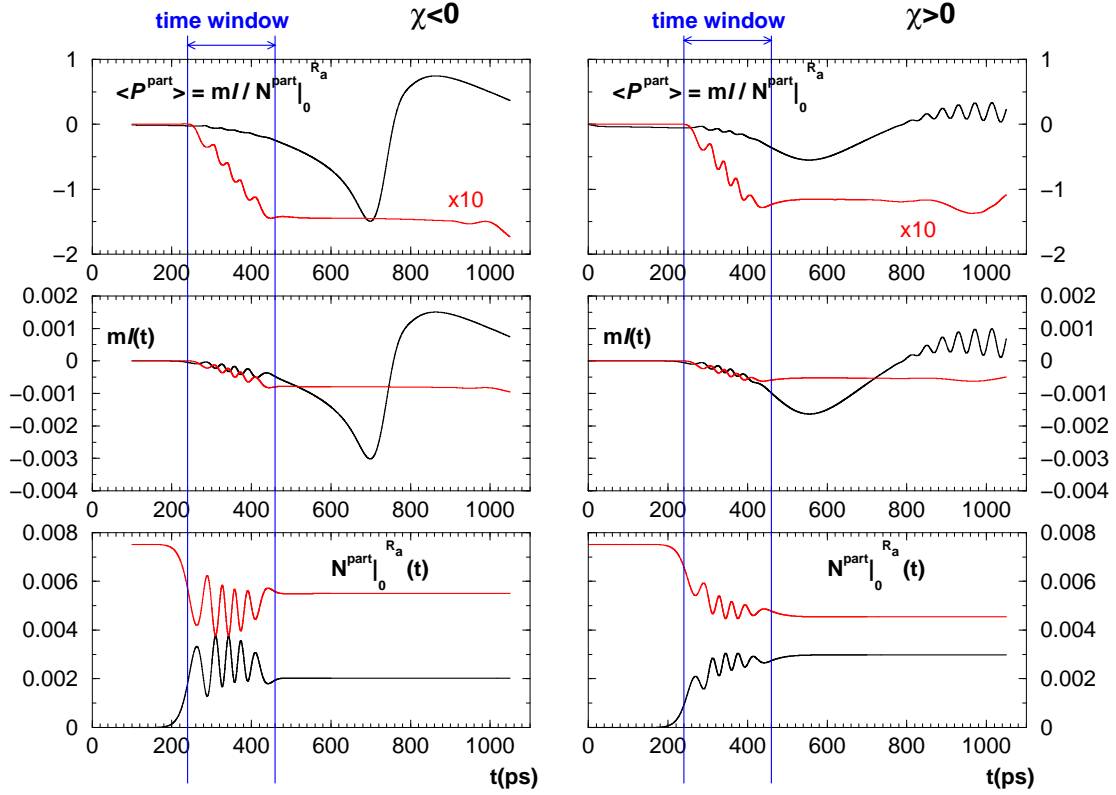}
\caption{Average momentum and influence of the sign of the chirp . Left pannel: Photoassociation with the pulse ${\cal{P}}^{122}_-$, with negative chirp : Comparison between expectation values of the momentum $<P^{\mathrm{part}}_g(t)>$ and the probability current $I_g(t)$ ,the population in the ground state $\mathrm{N}_{g}^{\mathrm{part}}(t)$ in the region R < 314 a$_0$(red lines) . The quantities for the excited state,  $<P^{\mathrm{part}}_e(t)>$ , $I_e(t)$ and $\mathrm{N}_{g}^{\mathrm{part}}(t)$ are drawn with black lines. . Right pannel: same quantities for photoassociation with the   pulse ${\cal{P}}^{122}_+ $, with positive chirp.}
\label{fig:vitessenegpos}
\end{figure}

The average value of the momentum gained during the pulse by the inner part of the ground state wavepacket can be evaluated from
\begin{equation}
<P^{\mathrm{part}}_g(t)>=\frac{I_{g}^{\mathrm{part}}(t)}{\mathrm{N}_{g}^{\mathrm{part}}(t)}
\end{equation}
This quantity is represented in the left upper pannel of \ref{fig:vitessenegpos}. The Rabi oscillations observed in the time-variation of $I_{g}^{\mathrm{part}}(t)$ and $\mathrm{N}_{g}^{\mathrm{part}}(t)$ being of opposite phase, they disappear almost completely in $<P^{\mathrm{part}}_g(t)>$, which proves again that the oscillations in $I_{g}^{\mathrm{part}}(t)$ are almost entirely due to the population. The mean negative velocity gained by the ground state wavepacket is $\sim$ 2.7 m s$^{-1}$, in agreement with the previously estimated value.
\subsection{Sensitivity to the sign of the chirp parameter and to the intensity}
The sensitivity to the sign of the chirp is illustrated in the right pannel of Fig. \ref{fig:vitessenegpos}, where are reported calculations done for the photoassociation  pulse ${\cal{P}}^{122}_+$ similar to the previous one but for $\chi>0$. The maximum population transferred to the excited state during the pulse is much smaller  than for a negative chirp, the amplitude of the oscillations being also smaller. However, after the pulse, the population remaining in the excited state is 0.003, larger by a factor 1.5 than for $\chi<0$. Looking at the mean value of the population, we see that recycling occurs for the pulse ${\cal{P}}^{122}_-$ whereas for a positive chirp a regular increase of the population is observed. This recycling effect can be attributed to the multiple interactions \cite{brown2006,wright2007} occurring for a negative chirp, when the motion of the two wavepackets is "following" the pulse, and not for a positive chirp.\\
After the ${\cal{P}}^{122}_+$ pulse, the wavepacket in the excited state is spreading, since there is no focussing effect : the reflexion on the inner wall takes place during a longer time interval than for $\chi<0$, from $t_P-\tau_C+\frac{1}{2}T_{vib}(v_{min})$= 560 ps to $t_P-\tau_C+\frac{1}{2}T_{vib}(v_{max})$=1000 ps. Starting from $t$= 800 ps, the large spead of the wavepacket gives rise to interferences between components moving in the opposite direction.\\
The mean momentum gained by the inner  part of the wavepacket after the ${\cal{P}}^{122}_+$ pulse  corresponds to a mean velocity of 2.1  m s$^{-1}$, a factor 1.3 smaller than in the case of the negative chirp.\\ 
It seems that the momentum kick is favoured by important Rabi oscillations and a smaller photoassociation efficiency.\\

\begin{figure}[htbp]
\includegraphics[width=0.9\textwidth]{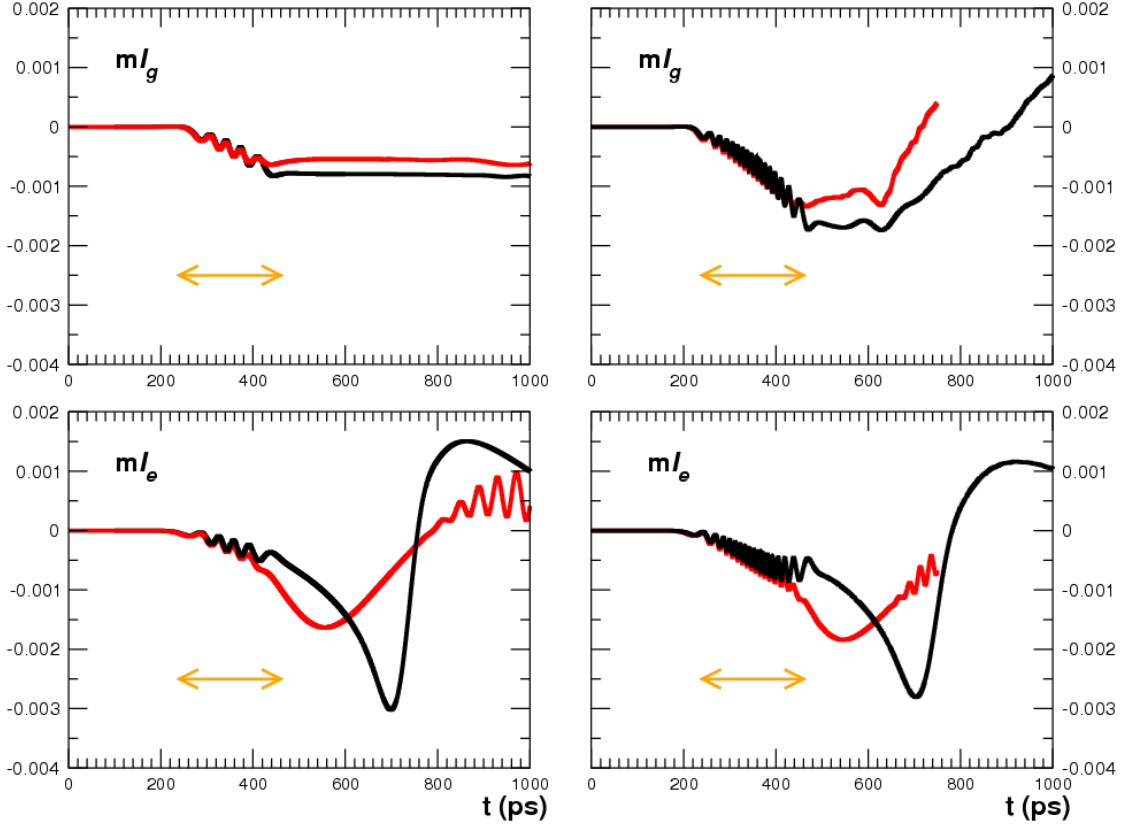}
\caption{(Color on line)Effect of the sign of the chirp and of the intensity of the photoassociation pulse.  Time-variationof the current $m I_g$ in the ground state (upper figures) and $m I_e$ in the excited state (lower figures)  . Left column, calculations with the pulses ${\cal{P}}^{122}_-$ (black lines), ${\cal{P}}^{122}_+$(red lines) as in Fig.\ref{fig:vitessenegpos}. Right column : same calculations, increasing the coupling in Eqs.(\ref{eq:eqcpl}) by a factor of 3, or the intensity $I_L$ in Eq.(\ref{eq:pulse}) by a factor of 9 and considering  negative (black lines) and positive (red lines) chirp. The increase of the momentum kick in the ground state is clearly  manifested}
\label{fig:compar3}
\end{figure}

The effect of the laser intensity is discussed in Fig. \ref{fig:compar3}. When the intensity is increased by a factor of 9, the number of Rabi oscillations during the pulse increases by a factor of 3 as expected. The transfer of population after the pulse $\mathrm{N}_{e}^{\mathrm{part}}(t>t+\tau_C)$ increases from 0.0024 for $\chi <0$ to 0.0038 for $\chi >0$. The population in the excited state is now only a factor of 1.15 larger for $\chi >0$. : this can be explained by the occurence, at larger intensities, of a recycling effect not only with negative but also with positive chirp.\\
When considering the larger intensity and a negative chirp,  the momentum kick in the inner part of the ground state wavepacket is increased by a factor of $\sim$ 2. The inner part of the wavepacket moves to short distances  with a mean velocity of $\sim$6 ms$^{-1}$ ( compared to 2.7 previously) : the reflexion on the inner wall of the \gts potential, with $I_P$=0, occurs $\sim$ 400 ps after the pulse maximum, to be compared with  the value 950 ps in the previous case. For positive chirp, the momentum kick in the ground state is now very close to the values obtained for $\chi <0$. However, since in the latter case the population $\mathrm{N}_{g}^{\mathrm{part}}(t>t+\tau_C)$ is larger by a factor of 1.5 than for $\chi >0$, a more important compression effect is to be expected for negative chirp.

The previous  analysis shows that the relative gain in momentum on the two channels can be modified by changing the parameters of the pulse. The way of optimizing the latter is an open problem.
\subsection{Local and non-local effects}
\begin{figure}[htbp]
\begin{center}
\includegraphics[angle=270,width=.9\textwidth]{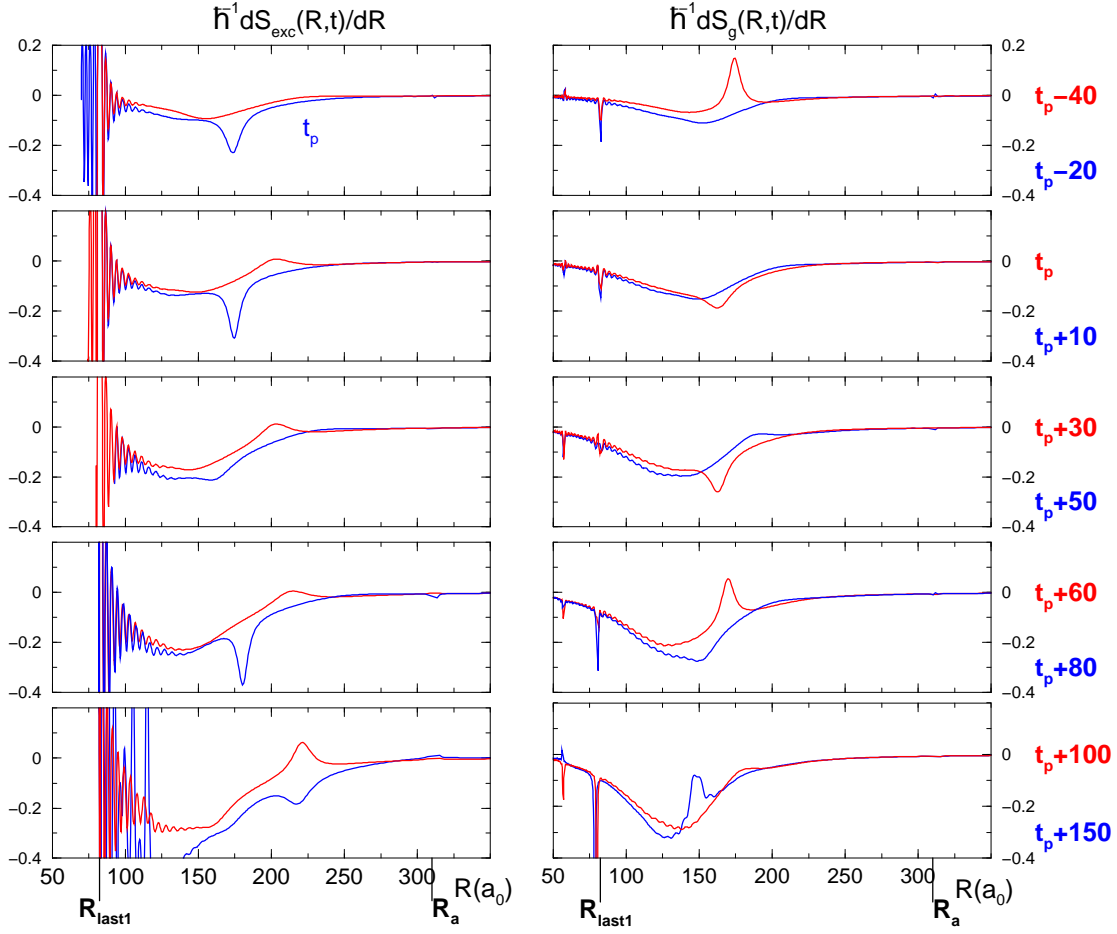}
\caption{(Color on line)Snapshots for various times during the pulse of variation of the local momentum $\frac{\partial d (S_{g/e}(R,t)}{\hbar\partial R}$, see Eq.(\ref{eq:phaseamp}) as a function of the internuclear distance $R$. This figure is giving more detailed information than Fig.\ref{fig:phase}; the strong variation of the phase as a function both of the distance $R$ and of the time $t$ is demonstrated. The position $R_a$ of the "last node" in the initial wavefunction, and $R_{last1}$ of the previous one are indicated by arrows;}
\label{fig:local}
\end{center}
\end{figure}

During the pulse, the variation of the local probability current density on each channel, as described in Eqs.(\ref{eq:accelg},\ref{eq:accele}), as well as the probability amplitudes (see Eqs. (\ref{eq:amplig},\ref{eq:amplie}) depend upon coupling terms involving the laser coupling $W_L(t) \exp(-i \phi(t))$ , and either $\Psi^*_e(R,t) \frac{\partial \Psi_g(R,t)}{\partial R}$ or $\Psi^*_e(R,t)$ $\Psi_g(R,t)$. The important quantity  is then the phase difference, 
\begin{equation}
\frac{S_g(R,t)}{\hbar}-\frac{S_e(R,t)}{\hbar}-\phi(t)
\label{phasedif}
\end{equation}
which depends both upon  $t$ and  $R$. In our problem, the initial wavefunction $\Psi_g(R,t)$ is delocalized, and extends over the full range of internuclear distances. Therefore during the pulse the excited wavepacket $\Psi_e(R,t)$  will also be delocalized. However, for very large distances, neither $S_g(R,t)$ nor  $S_e(R,t)$ present a $R $ dependence. In  Fig. \ref{fig:local}, the $R$-variation of $\frac{\partial  (S_{g/e}(R,t)}{\hbar\partial R}$ is represented for various times in the vicinity of the pulse maximum. For distances up to 250 a$_0$, very rapid changes  of the phase occur, and a very complicated $R$-dependence is manifested. 
 manifested.
This suggests that the choice of $R$-integrated physical quantities, as used in Refs. \cite{brown2006, kallush2007b}, to optimize the pulse, may not be a very efficient procedure when the delocalized character of the initial wavefunction is introduced in the model. \\
 In the "single pulse Raman-like walk scheme" of Ref.\cite{brown2006}, the initial state is a localized gaussian wavepacket, and the control parameter is the laser instantaneous frequency, designed to match the collision dynamics at distances $R<$40 a$_0$, by inducing multiple coherent upward and downward transitions. It seems that for this "matched spectrum pulse", a lot of population is redistributed into continuum levels of the ground state, creating pairs of hot atoms rather than the target bound levels in the ground state. In Ref.\cite{kallush2007b}, the global phase of  matrix elements such as $<\Psi_e(R,t)|D(R)|\Psi_g(R,t)>$ and $<\Psi_e(R,t)|D(R)|\frac{\partial\Psi_g(R,t)}{\partial R}>$ is the control parameter. The phase of the field is then following the variation in the phase of the expectation value to be controlled. It then seems that for pulse duration larger than 1 ps , the efficiency of the control is dropping.\\
 The choice of a spatially global operator such as the total population
or the total momentum can overlook the finer details of controling a localized quantity.
This is equivalent to the $R$-integrated quantities in the present paper, which loose a large part of the phase information. 
Therefore, the partially integrated quantities (\ref{eqn:partgen},\ref{eqn:part}) are  a better choice, 
enabling a better  optimization of the compression effect.
In the next paper \cite{luc2007b} the current approach will be linked to a different set of  spatially semilocal operators.\\
However, from the present analysis of the partially integrated quantities, we have drawn a few conclusions that might help to optimize the compression effect.

\section{Control of the compression effect}
\label{sec:control}
\begin{figure}
\includegraphics[width=.7\textwidth]{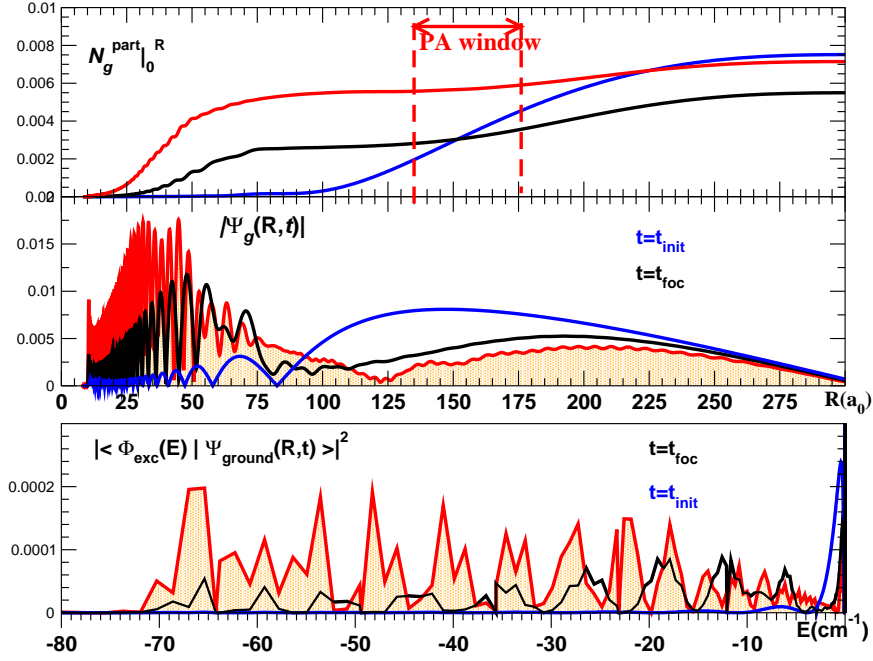}
\caption{Improvement of the compression effect presented in Fig. \ref{fig:overlap}, using the new pulse described in Section \ref{sec:control}, with a parameter $\tau_C$ comparable to the half vibration period in the excited state. Intermediate panel : wavefunction $|\Psi_g(R,t=t_{\mathrm{init}})|$  of the initial collision state and ground state wavepacket $|\Psi_g(R,t_{\mathrm{opt}}|$ when the compression effect is maximum, after illuminating with the pulse ${\cal{P}}^{122}_0$ (black line,$t_{\mathrm{opt}}=t_P$+ 950 ps) and the new pulse (red line, $t_{\mathrm{opt}}=t_P$+ 350 ps). Upper panel: integrated population $\int_0^R |\Psi_g(R,t)|^2$, showing the increase of the probability density at short distances. Lower panel : variation of the overlap $|<\Phi_e(v)|\Psi_g(R,t)>|^2$  with the eigenfunctions for all the  bound vibrational levels $v$ in  the $0_g^-(6S+6P_{3/2}$) excited potential curve, as a function  of their  binding energy $E_B(v)$, for $t=t_{\mathrm{init}}$(blue) and $t=t_{\mathrm{opt}}$(black and red line).  The compressed wavepacket is even better adapted for further photoassociation.}
\label{fig:reexcit}
\end{figure}

From the discussion of Section \ref{sec:analysis}, it seems that the creation of an important flux of population to short distances is favoured both by an important   population cycling during the pulse and by a good transfer of this population back to the ground state. In other words, the optimal pulse should be as close as possible to a $(2n+1)\Pi$ pulse, with a negative chirp, and a large number $n$ of cycles. This can be achieved with a pulse that is "following" the excited wave packet.
 We have tentatively changed the parameters of the photoassociation pulse to check whether it is possible to optimize the compression effect. Starting from a transform limited pulse with the same spectral width as previously ( $\delta \omega$=.255 \cm, $\tau_L$= 57.5 ps) and with the same central frequency, the best results were obtained by increasing the linear chirp parameter $\chi$ and hence the temporal width. The    optimal value was $\tau_C$ = 376.13 ps, instead of 110 ps in the previous calculations. The duration of the pulse then becomes comparable with half the vibration period ($T_{vib}/2$)in the excited state. In order to make results comparable, the $\epsilon_0$ factor in \ref{eq:W} was mutiplied by a factor  compensating the variation of$\sqrt{\tau_L/\tau_C}$ factor in $f(t)$ in order to keep constant the maximum value of the coupling. The energy  $E_{pulse}$ is then multiplied by 6.25. The new pulse is closer to a $(2n+1)\Pi$ one, the number of Rabi oscillations reaches up to 24, the population transferred to the excited state has markedly decreased (3 10$^{-4}$ instead of 2 10$^{-3}$). \\
The compression effect is now much more important, occurs earlier and is optimum at $t_{opt}=t_P$+~350~ps. The wavepacket at this time is displayed  in Fig.\ref{fig:reexcit}, demonstrating a spectacular increase of the compression effect as compared to the previous results, and suggesting that the direction is the good one.\\
 
The analysis of the results shows that the model of adiabatic transfer within a photoassociation window is still valid, but the impulsive approximation is no longer valid; new physical effects are present, which will be analyzed in a forthcoming paper \cite{luc2007b}.
\section{Conclusion}
The present paper was devoted to the analysis of the formation of a dynamical hole in the   wavepacket describing the relative motion of a pair of cold atoms in presence of a photoassociation pulse,  and of the evolution of this wavepacket after the pulse. As a case study, photoassociation of Cs into loosely bound levels of the outer well in  Cs$_2$\zerog was considered, using a pulse in the picosecond range,  with a negative linear chirp parameter, designed in previous work \cite{luc2004a,koch2006b} to create an excited wavepacket with focussing properties. In the initial state, the wavepacket is a stationary continuum wavefunction at the collision energy corresponding to 54 $\mu$K. Photoassociation with such a pulse can be interpreted as  an adiabatic transfer of population within a photoassociation window, while the impulsive approximation is valid, since the motion of the nuclei can be neglected during the pulse. The numerical calculations show the creation of a dynamical hole in the region of the photoassociation window. The latter is created at distances where the ground state potential is negligible, while the excited state potential, with asymptotic $R^{-3}$ behaviour, is giving to the excited wavepacket an acceleration towards short distances. \\
After the pulse, the inner part of the ground state wavepacket is shown to move towards short distances, at a velocity typical of the vibrational motion in the excited state, due to the negative momentum kick gained from interaction with the laser field. This leads to a compression of population at short range, so that a second photoassociation pulse, red-detuned relative to the first one, and conveniently delayed, could be designed to bring much population to deeply bound levels of the excited state. Another signature of the dynamical hole is the redistribution of population in the ground state, with formation of pairs of hot atoms and population of the last bound level.\\
We have analyzed the momentum kick phenomenon by considering the time-evolution of the position probability density and of the probability current density. Spatially integrated values, giving the population and momentum on each channel, were considered, as well as partially integrated values, considering population below a certain distance. This procedure is necessary due to the delocalized character of the initial wavepacket. The various terms included in the treatment have been discussed in the framework of a phase amplitude formalism: besides the classical force and the force due to the laser, a third term has been pointed out. \\
The analysis for a pulse in the 100 ps range, widely used in previous work, with negative chirp, shows that, during the pulse duration, population and momentum exchange between the two channels occur due to Rabi cycling, so that the acceleration from the excited channel is transferred to the ground channel. After the pulse, the momentum gained by the inner part of the ground state wavepacket depends upon the number of Rabi cycles, upon the sign of the chirp ( it is  smaller in case of a positive chirp) and upon the amount of population transferred back to the ground state. It seems to be maximum for a pulse as close as possible to a (2n+1) $\Pi$ pulse, with which no photoassociated molecules are formed. In contrast, when considering a similar pulse with positive chirp, the photoassociation efficiency increases while the momentum kick is reduced. \\
Following these ideas, we have tentatively proposed photoassociation by a longer negatively chirped pulse, with duration similar to half the classical vibrational period in the excited state, so that the impulsive approximation is no longer valid. Such a pulse is designed to " follow" the motion of the excited wavepacket during half the vibrational period. A strong increase of the momentum kick and of the compression effect in the ground state was indeed obtained, demonstrating that the direction is promising. Further analysis will be given in a forthcoming paper.  
\begin{acknowledgments}
  
  This work has been supported by the European Union in the frame of the 
  Cold Molecule EC network under contract HPRN-CT-2002-00290. Ronnie Kosloff acknowledges for a one month invitation from Université Paris-Sud XI as "professeur invit\'e" in Laboratoire Aim\'e Cotton. 
  The Fritz Haber Center is supported
  by the Minerva Gesellschaft f\"{u}r die Forschung GmbH M\"{u}nchen, Germany. Laboratoire Aim\'e Cotton is part of F\'ed\'eration Lumi\`ere Mati\`ere (LUMAT, FR 2764). 
\end{acknowledgments}

\end{document}